\pdfoutput=1

\documentclass[10pt,a4paper,twoside]{baltlat7}
\usepackage{epsfig}
\begin{document}
\ \
\vspace{0.5mm}

\setcounter{page}{205}
\vspace{8mm}

\titlehead{Baltic Astronomy, vol.\,18, 205--215, 2009}

\titleb{CHAOS IN A DISK GALAXY MODEL INDUCED BY \\ ASYMMETRIES IN THE
DARK HALO}

\begin{authorl}
\authorb{N. D. Caranicolas}{} and
\authorb{E. E. Zotos}{}
\end{authorl}

\moveright-3.2mm
\vbox{
\begin{addressl}
\addressb{}{Department of Physics, Section of Astrophysics, \\
 Astronomy and Mechanics, Aristotle University of Thessaloniki, \\
 541 24  Thessaloniki,  Greece; caranic@astro.auth.gr}
\end{addressl}}

\submitb{Received: 2009 June 10; accepted: 2009 September 22}

\begin{summary} We study the regular or chaotic nature of motion in a
disk galaxy with a dense nucleus and an asymmetric dark halo.  Two
cases, the 2D model and the 3D model, are investigated.  In the 2D
model, a considerable fraction of the phase plane is covered by chaotic
orbits.  Two factors seem to be responsible for the chaotic motion:  (i)
the dense nucleus and (ii) the asymmetries in the dark halo.  Our
numerical experiments suggest, that there are several chaotic components
on the Poincar\'e phase plane.  Different chaotic components are
induced by the asymmetries in the halo.  Each chaotic component seems to
have a different value of the Lyapunov Characteristic Exponent, for
small values of the asymmetry parameter $\lambda$ and a unique LCE for
larger values of $\lambda$.  A comparison of the present results with
outcomes from previous work is also presented.  \end{summary}


\resthead{Chaos in a disk galaxy model induced by dark halo}{N. D.
Caranicolas, E. E. Zotos}

\sectionb{1}{INTRODUCTION}

Galaxies are often surrounded by halos.  In most cases the shape of
halo is spherical or nearly spherical but there are also indications
that the shape of halo may be a biaxial or even a triaxial ellipsoid
(see Kunihito et al. 2000; Olling \& Merrifield 2000; Wechsler et al.
2002; Caranicolas \& Zotos 2009).  The distinction between the halo and
the main body of the galaxy is clearest in disk galaxies, where the
spherical or the triaxial shape of the halo contrasts with the flat
disc.  In an elliptical galaxy, there is no sharp transition between the
body of the galaxy and the halo.  The visible part of the halo, is
occupied by population II objects, including globular clusters and old
individual stars.  Beyond this, is a much larger region, called the dark
halo or extended halo, containing large amounts of dark matter.  Note
that the stellar halo is not an inner part of the dark halo.  They are
two physically distinct components with different formation histories.
We would like to point out that the above mentioned triaxial dark halo
models are all symmetrical, unlike the dark halo model in the present
paper.

The need for existence of dark halos results from many independent
sources such as rotation curves, hot gas in clusters, velocity
dispersions of galactic groups, gravitational lensing and microwave
radiation.  The presence of dark matter in the halo is demonstrated by
its gravitational effect on the rotation curve of the disk.  Without
large amounts of mass in the extended halo, the rotational velocity of
the galaxy should decrease at large distance from the galactic core.
However, observations of disk galaxies, particularly radio observations
of line emission from neutral atomic hydrogen, show that the rotation
curve of most spiral galaxies remains flat far beyond the visible
matter.  The absence of any visible matter to account for these
observations implies the presence of unobserved matter.  The nature of
dark matter in the galactic halo of spiral galaxies is still
undetermined, but the most popular theories accept that the dark halo is
populated by vast numbers of small bodies known as MACHOs.  Observations
of the halo of the Milky Way show that the number of MACHOs is not
likely to be sufficient to account for the required mass.

No doubt that the behavior of orbits in the galactic disk is affected by
the presence of the halo.  Of particular interest is to study the effect
of the asymmetries of the dark halo in the behavior of orbits in the
disk.  In order to investigate this behavior we have constructed a
composite 3D dynamical model of a disk galaxy.  The model is described
in Section 2. In Section 3 we use the $x-p_x$ phase plane in order to
study the motion in the 2D model.  In Section 4 we use the results
obtained from the 2D model in order to visualize the motion in the 3D
model.  Of special interest is to find the behavior of 3D orbits in
different chaotic components observed in the 2D model.  We close our
research with discussion and conclusions which are presented in
Section 5.

\sectionb{2}{PRESENTATION OF THE DYNAMICAL MODEL}

Our model consists of three distinct components -- the disk halo, the
nucleus and the dark halo component.  The disk halo component is
described by the potential
\begin{equation}
V_d(x,y,z)=\frac{-M_d}{\sqrt{b^2+x^2+y^2+(a+\sqrt{h^2+z^2})^2}},
\end{equation}
where $M_d$ is the mass, $b$ is the core radius of the disk halo,
$\alpha$ is the disk scale length and $h$ is the disk scale height.
Potential (1) was introduced by Miyamoto \& Nagai (1975).  The dense
massive nucleus is described by the spherical potential
\begin{equation}
V_n(x,y,z)=\frac{-M_n}{\sqrt{x^2+y^2+z^2+c_n^2}},
\end{equation}
where $M_n$ is the mass and $c_n$ is the scale length of the nucleus.
For the dark halo we use the logarithmic potential
\begin{equation}
V_h(x,y,z)=\frac{\upsilon_0^2}{2}ln[x^2+y^2+z^2-\lambda x^3+c_h^2].
\end{equation}
Here $c_h$ is the scale length of the dark halo component and the
parameter $\upsilon_0$ is used for the consistency of the galactic
units.  The term $-\lambda x^3, \lambda \ll 1$ represents the deviation
of the halo from spherical symmetry (see also Binney \& Tremaine 2008).

We use the system of galactic units, where the unit of length is 1 kpc,
the unit of mass is $2.325 \times 10^7 M_\odot$ and the unit of time is
$0.97748 \times 10^8$ yr.  The velocity unit is 10 km/s, while $G$ is
equal to unity.  The energy unit (per unit mass) is 100 (km/s)$^2$.  In
the above units we use the values:  $\upsilon_0=20, \alpha=3, b=6,
h=0.325, M_d=12000$.  The value of $c_n$ and $c_h$ is 0.25 and 8
respectively while $M_n$ and $\lambda$ are treated as parameters.

The total potential responsible for the motion of a test particle of
unit mass in the galaxy is
\begin{equation}
V_T(x,y,z)=V_d+V_n+V_h.
\end{equation}

There are three lines of arguments to justify the choice of potential
(4).  The first is that the disk nucleus potential system given by
equations (1) and (2) is a classical realistic potential, which
describes very well the motion in an active disk galaxy.  The second is
that potential (3) describes satisfactory the deviation of the dark
halo from spherical symmetry, and the third is that there is
observational evidence that asymmetries in the halos do exist (see Xu et
al. 2007).  However we must emphasize that (Xu et al. 2007) investigated
the stellar halos, not the dark halos.

The equations of motion are:
\begin{equation}
\ddot{x}=-\frac{\partial V_{T}}{\partial x},
\ddot{y}=-\frac{\partial V_{T}}{\partial y},
\ddot{z}=-\frac{\partial V_{T}}{\partial z},
\end{equation}
where the dot indicates derivative with respect to the time.  The
corresponding Hamiltonian is
\begin{equation}
H=\frac{1}{2}(p_x^2+p_y^2+p_z^2)+V_T(x,y,z)=E,
\end{equation}
where $p_x, p_y , p_z$ are the momenta per unit mass, conjugate to $x$,
$y$ and $z$ while $E$ is the numerical value of the Hamiltonian.

\sectionb{3}{NUMERICAL RESULTS FOR THE 2D MODEL}

In this case we integrate numerically the equations of motion (5) with
$z=p_z=0$.  The corresponding Hamiltonian is
\begin{equation}
H_2=\frac{1}{2}(p_x^2+p_y^2)+V_T(x,y)=h_2,
\end{equation}
where $h_2$ is the numerical value of the energy of the test particle.

We believe that it is a good idea to start from the 2D system for two
basic reasons:  (i) in the 2D system we can use the $x-p_x, y=0, p_y>0$
Poincar\'e phase plane in order to locate the areas of regular and
chaotic motion, and (ii) we can use the experience gained from the study
of the 2D system in order to explore a more complicated 3D system.

Figure 1 shows the $x-p_x$ phase plane when $\lambda=0.015, M_n=400,
h_2=500$.  We see that the larger part of the phase plane is covered by
regular orbits.  There are also parts of the phase plane occupied by
chaotic orbits.  Three distinct areas of chaotic motion are observed:  a
considerable chaotic layer near the central region and two secondary
chaotic components which appear near the unstable periodic orbit
produced by secondary resonances.  Figure 2 is similar to Figure 1 but
for $h_2=300$.  Here one can see only the chaotic layer near the central
part of the phase plane while no other chaotic component seems to be
present.  The rest of the phase plane is covered by regular orbits.
Thus we may conclude that, as a result, the asymmetry in the dark halo
has to produce a different chaotic component for the high energy stars.
In other words, the asymmetry affects not only stars approaching the
dense nucleus but also stars moving far from the nuclear region.  On the
other hand, from the totality of low energy stars only those approaching
the nucleus are on chaotic orbits.

\begin{figure}[!th]
\resizebox{\hsize}{!}{\rotatebox{0}{\includegraphics*{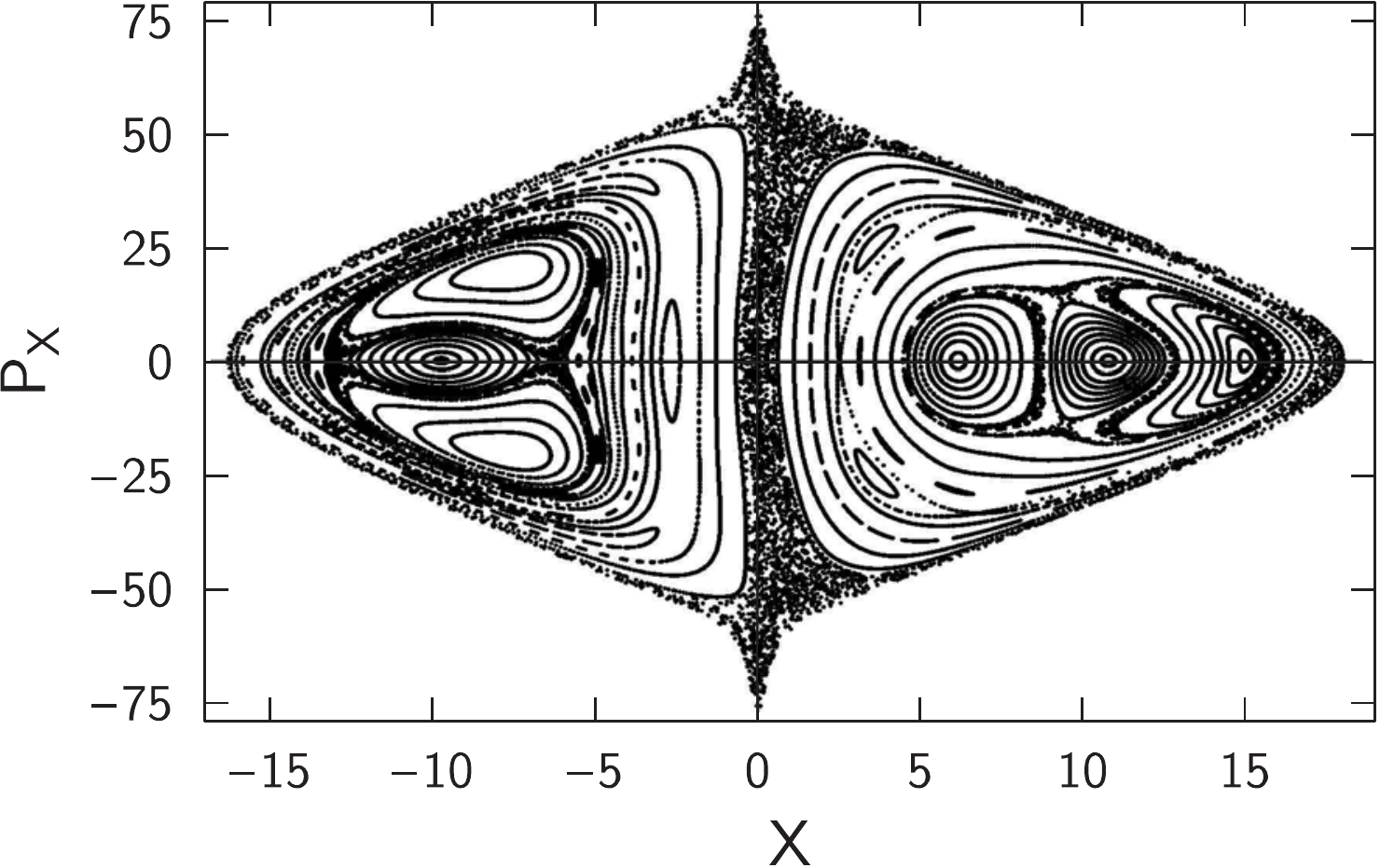}}}
\vskip2mm
 \captionb{1}{The $x-p_x$ phase plane of the 2D system. The values of
 the parameters are: $\upsilon_0=20, \alpha=3, b=6, h=0.325, M_d=12000,
c_n=0.25, c_h=8, M_n=400$, $\lambda=0.015$ and $h_2=500$.}
%\end{figure}
\vskip7mm
%\begin{figure}[!th]
\resizebox{\hsize}{!}{\rotatebox{0}{\includegraphics*{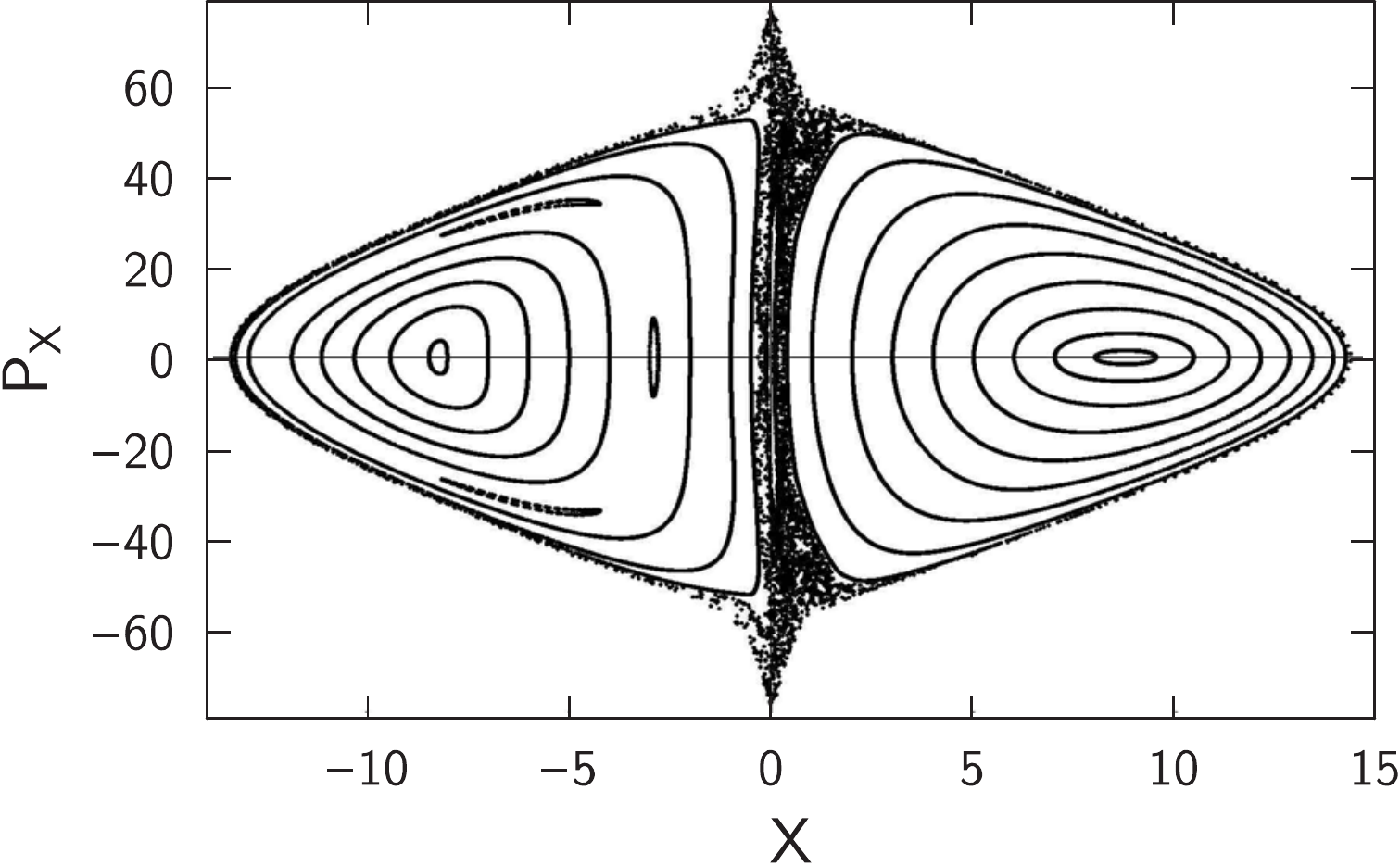}}}
\vskip2mm
\captionc{2}{The same as in Figure 1 but for $h_2=300$.}
\end{figure}

\begin{figure}[!th]
\resizebox{\hsize}{!}{\rotatebox{0}{\includegraphics*{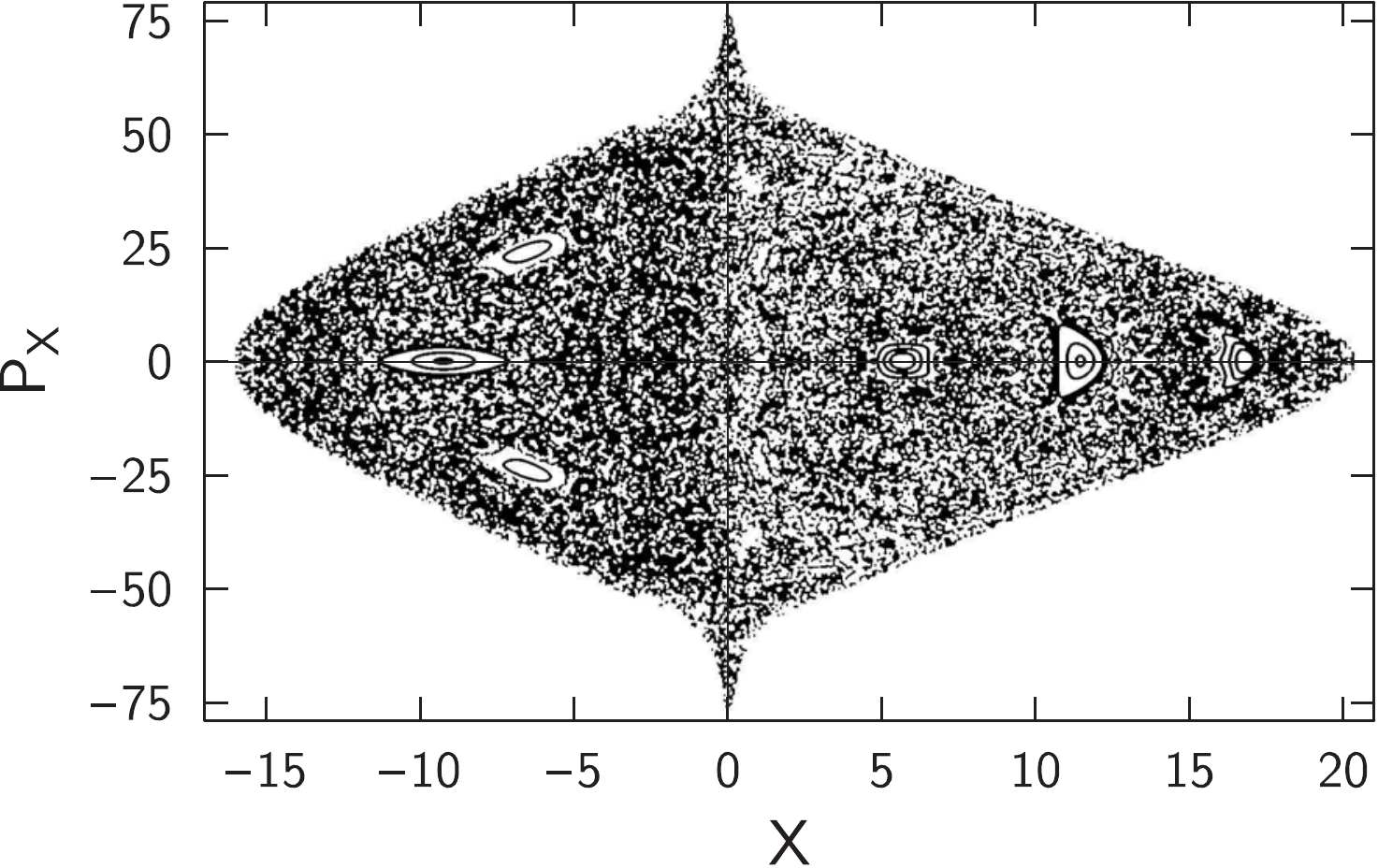}}}
\vskip2mm
\captionb{3}{The $x-p_x$ phase plane of the 2D system. The values of
the parameters are: $\upsilon_0=20, \alpha=3,  b=6, h=0.325, M_d=12000,
c_n=0.25, c_h=8, M_n=400$, $\lambda=0.03$ and $h_2=500$.}
%\end{figure}
\vskip7mm
%\begin{figure}[!th]
\resizebox{\hsize}{!}{\rotatebox{0}{\includegraphics*{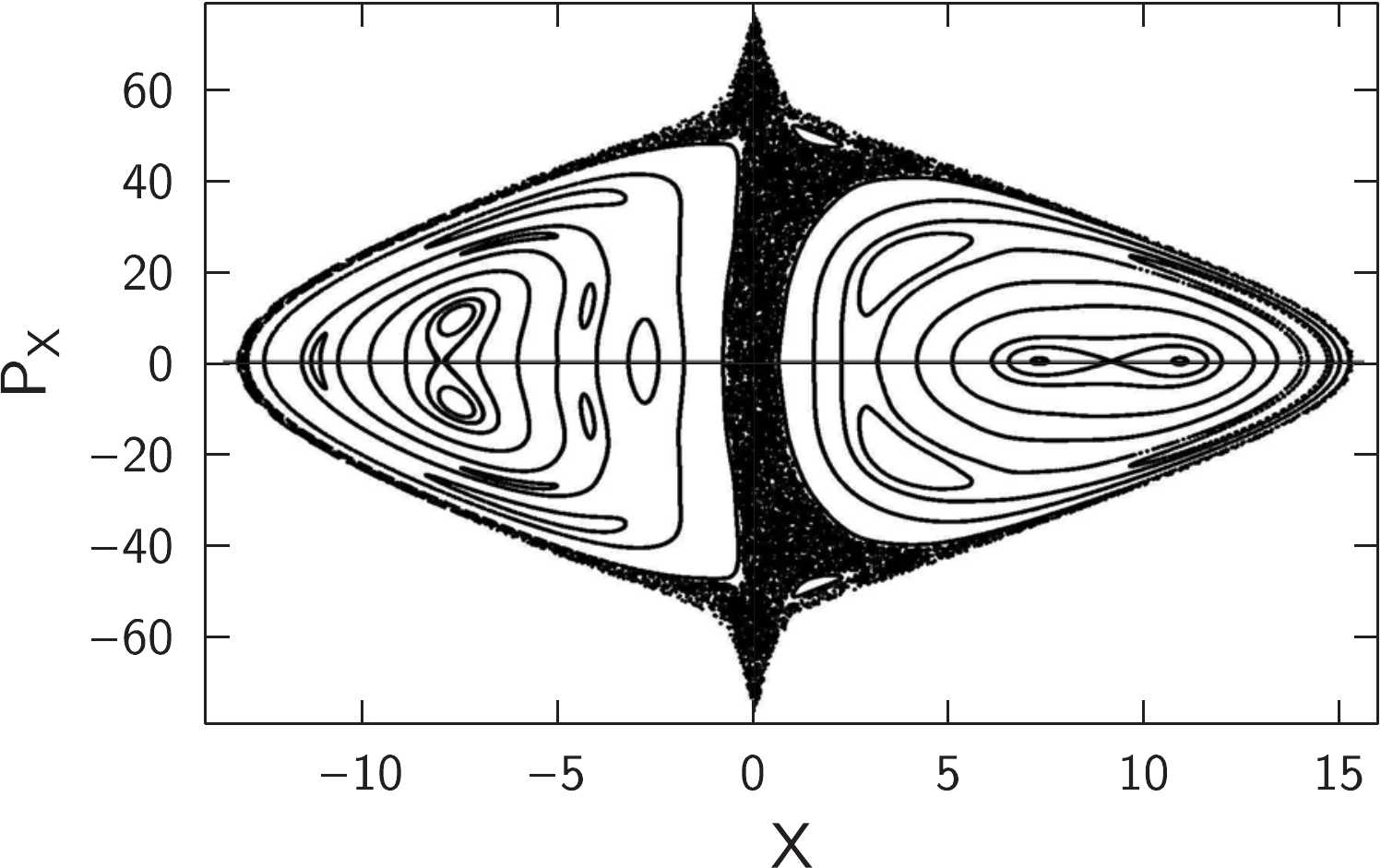}}}
\vskip2mm
\captionc{4}{The same as in Figure 3 but for $h_2=300$.}
\vspace{-.2mm}
\end{figure}

Figure 3 is similar to Figure 1 but for $\lambda=0.03$.  In Figure 3 we
see that the three chaotic components have merge to produce a large
chaotic sea, while the small regular regions are confined near the
stable resonant periodic points.  Furthermore, in Figure 4 we see that
the picture is almost the same as that shown in Figure 2. The only
difference is that here one can observe some additional small islands
produced by secondary resonances.  Thus, one can see that the increase
of the asymmetry in the halo affects drastically the high energy stars,
producing large chaotic regions by merging different chaotic components.
Moreover, low energy stars seem to display only one chaotic layer, which
is produced by low energy stars approaching the dense nucleus.

Figures 5(a--d) show four typical orbits for the 2D potential.  The
values of the parameters are as in Figure 1. Panel (a) shows a quasi
periodic orbit.  The initial conditions are:  $x_0=-6.5, y_0=0,
p_{x0}=19, h_2=500$.  The value of $p_y$ is always found from energy
integral (7).  Panel (b) shows a tube orbit with the initial conditions
$x_0$ = 7.0, $y_0$ = 0, $p_{x0}$ = 0 and $h_2$ = 300.  A quasi periodic
orbit is also presented in panel (c).  The initial conditions are:
$x_0$ = --3.0, $y_0$ = $p_{x0}$ = 0 and $h_2$ = 300.  Note that all the
above quasi periodic orbits do not approach the dense nucleus.  Panel
(d) shows a chaotic orbit with the initial conditions $x_0$ = 0.5, $y_0$
= $p_{x0}$ = 0 and $h_2$ = 500.  This orbit comes arbitrary close to the
nucleus.  The orbits were calculated for a time period of 100 time
units.

\begin{figure}[!ht]
\resizebox{\hsize}{!}{\rotatebox{0}{\includegraphics*{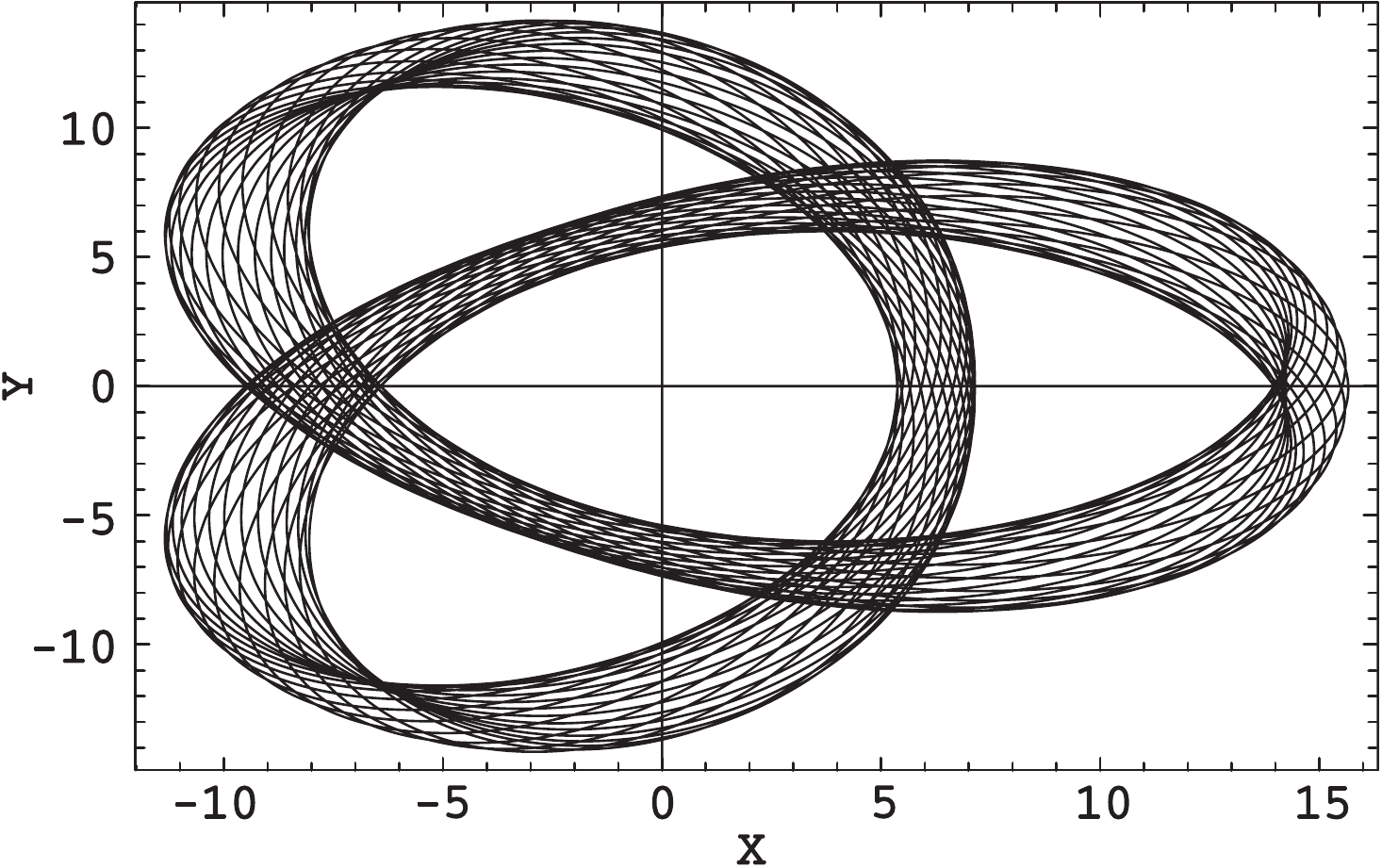}}\hspace{1cm}
                      \rotatebox{0}{\includegraphics*{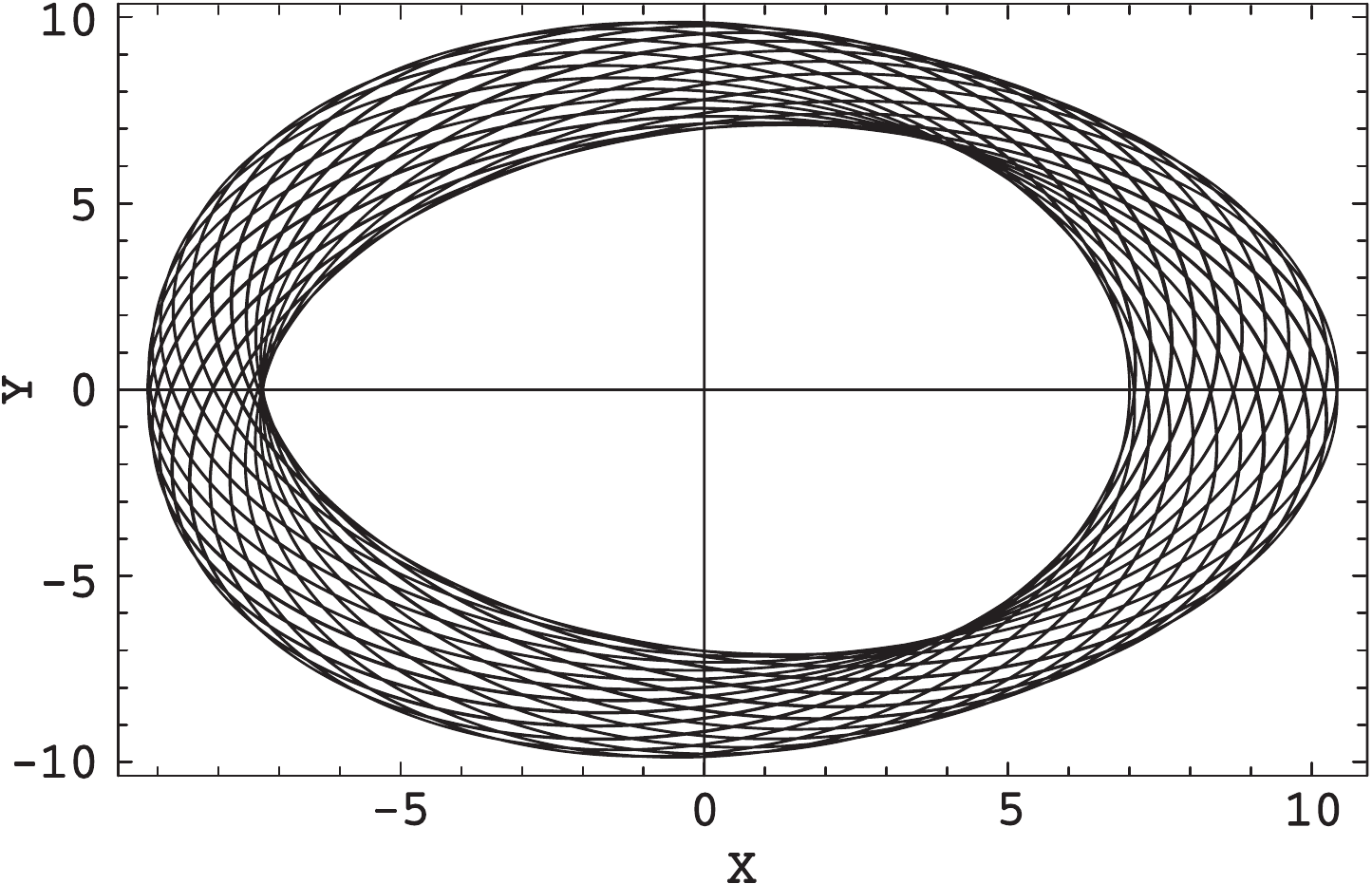}}}
\resizebox{\hsize}{!}{\rotatebox{0}{\includegraphics*{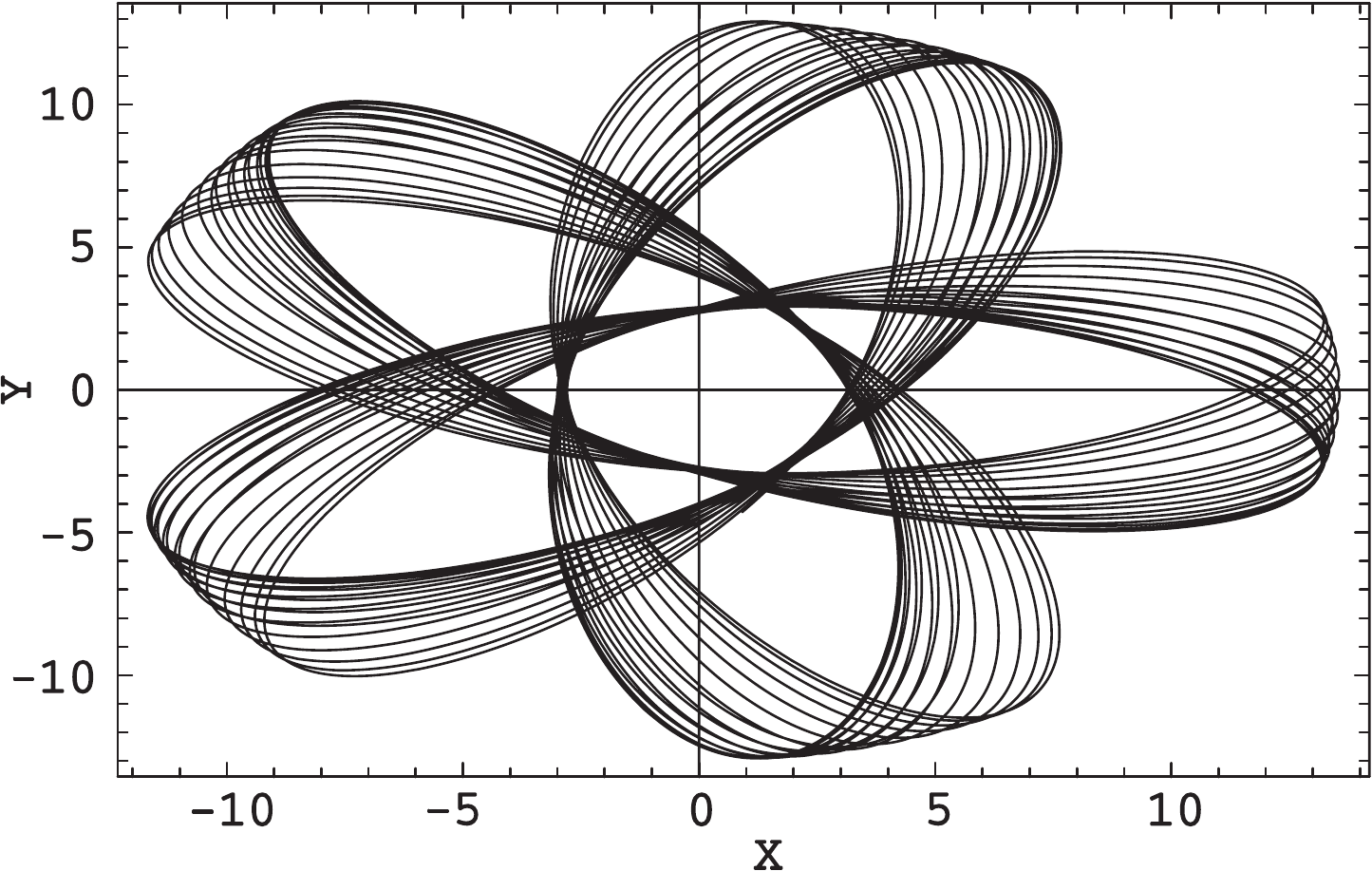}}\hspace{1cm}
                      \rotatebox{0}{\includegraphics*{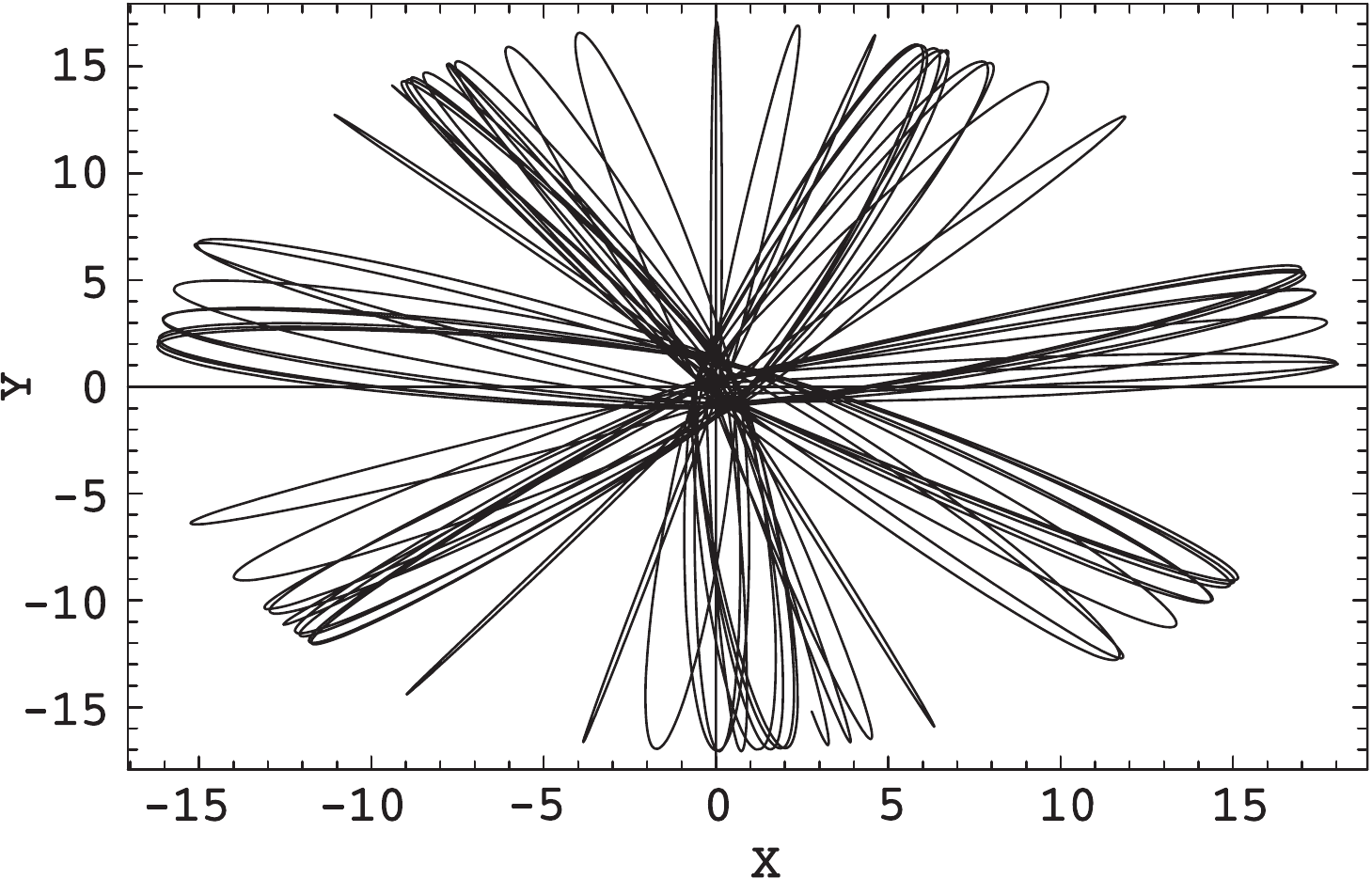}}}
\vskip 0.1cm

\captionb{5}{Panels (a)--(d):  typical orbits for the 2D potential.  The
values of the parameters are as in Figure 1. Panel (a) -- upper left:  a
quasi-periodic orbit with the initial conditions:  $x_0=-6.5$, $y_0=0$,
$p_{x0}=19$ and $h_2=500$.  Panel (b) -- upper right:  a tube orbit with
the initial conditions:  $x_0=7.0$, $y_0=0$, $p_{x0}=0$ and $h_2=300$.
Panel (c) -- lower left:  a quasi periodic orbit with the initial
conditions $x_0=-3.0$, $y_0=0$, $p_{x0}=0$ and $h_2=300$.  Panel (d) --
lower right:  a chaotic orbit with the initial conditions $x_0=0.5$,
$y_0=0$, $p_{x0}=0$ and $h_2=500$.  The value of $p_y$ is always found
from the energy integral.}

\end{figure}

It would be of particular interest to follow the evolution of chaotic
regions, as the parameter $\lambda$ increases or as the mass of the
nucleus $M_n$ increases, when all other parameters are kept constant.
The results are presented in Figures 6 and 7.  Figure 6 shows a plot
of the percentage of the surface of section $A\%$ covered by chaotic
orbits vs.  $\lambda$, when $M_n=400$.  The values of all other
parameters are as in Figure 1. Dots indicate the values found
numerically, while the solid line is a two degree polynomial fit.  We
see that $A\%$ increases rapidly as $\lambda$ increases.  Figure 7
shows a plot of $A\%$ vs.  $M_n$, when $\lambda=0.015$.  The values of
all other parameters are as in Figure 1. Again, dots indicate the values
found numerically, while the solid line is a two degree polynomial fit.
Here we see that $A\%$ increases as $M_n$ increases but at a smaller
rate.

\vskip3mm
%%%%%%%%%%%%%%%%%%%%%  FIGURES 6 and 7
\begin{figure}[!th]
\vbox{
\centerline{\psfig{figure=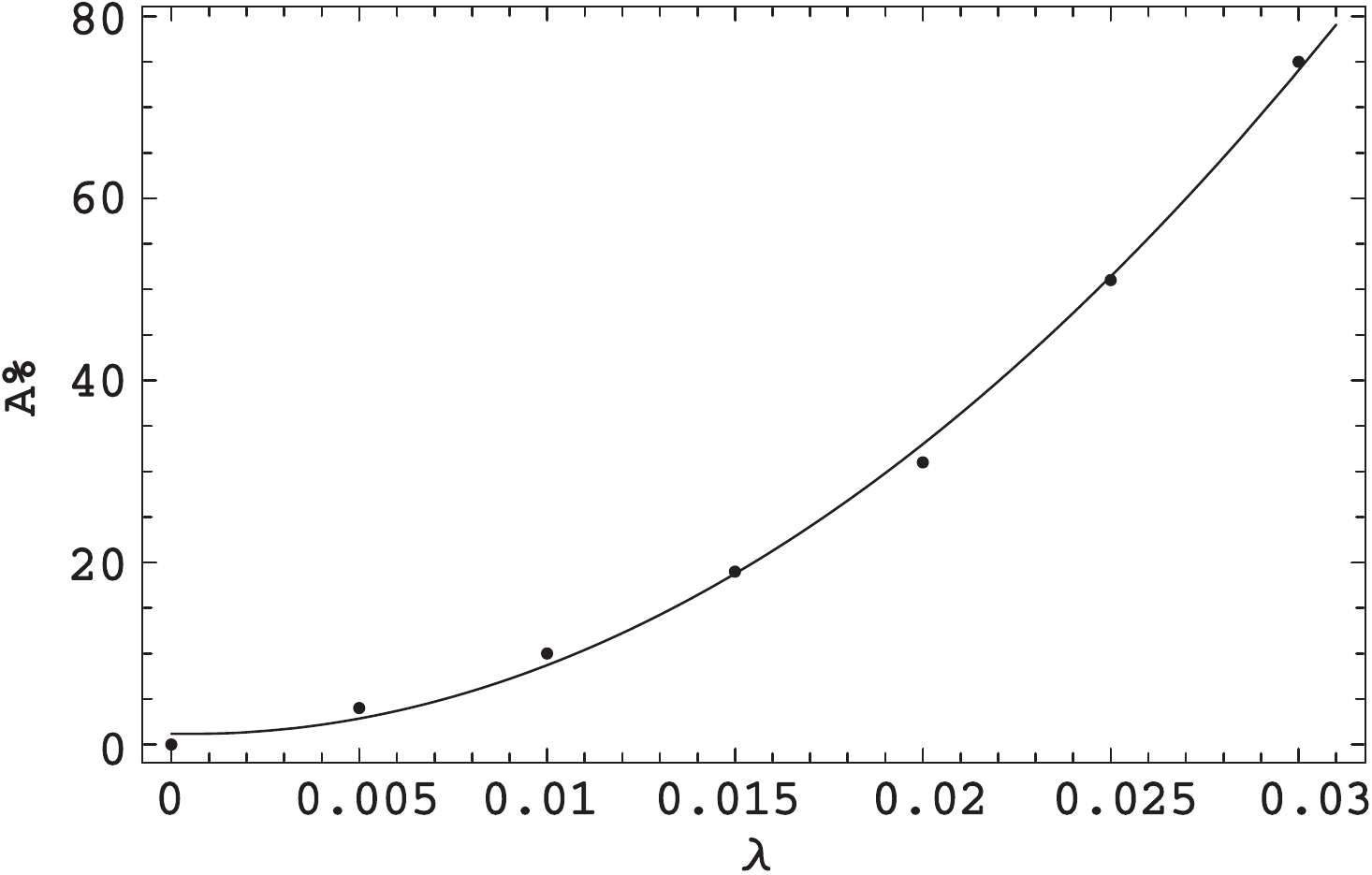,width=95mm,angle=0,clip=}}
\vspace{0.5mm}
 \captionb{6}{A plot of the percentage of the surface of section $A\%$
covered by chaotic orbits vs. $\lambda$, when $M_n=400$.}
\vspace{3mm}
\centerline{\psfig{figure=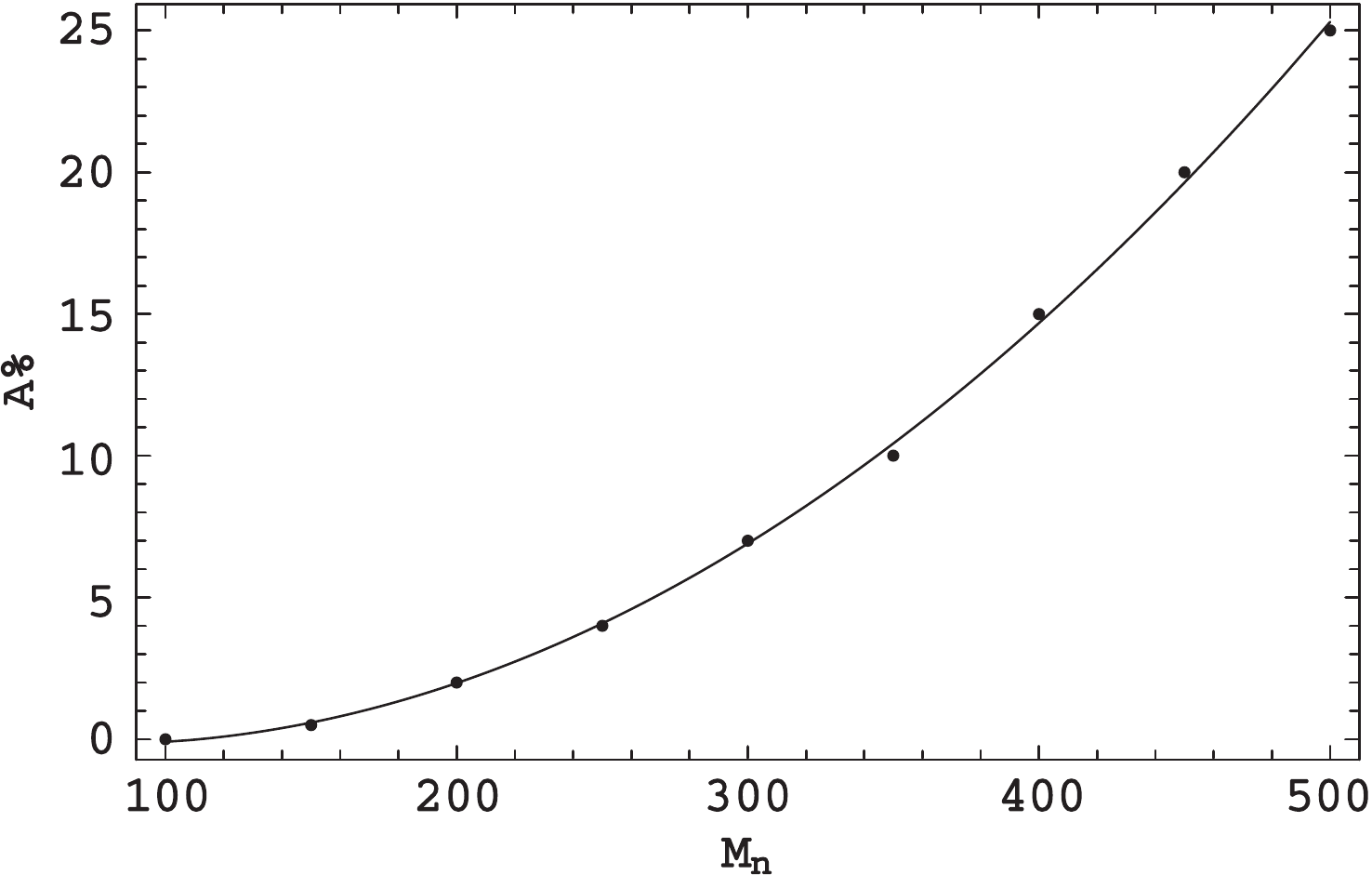,width=95mm,angle=0,clip=}}
\vspace{0.5mm}
\captionb{7}{A plot of the  percentage of the surface of section $A\%$
covered by chaotic orbits vs. $M_n$, when $\lambda=0.015$.}
}
\end{figure}
%\vskip5mm
%\newpage

It is also of interest to compute the Lyapunov Characteristic Exponents
(LCE) (see Lichtenberg \& Lieberman 1992) in order to estimate the
degree of chaos in different chaotic regions of Figure 1, i.e., when
$\lambda=0.015, M_n =400, h_2=500$.  The LCE in all cases was computed
for a time period of 20000 time units.  Note that this time period is at
least 100 times larger than the corresponding Hubble time.  The LCE was
found to be 0.195 in the chaotic region near the center.  The value of
the LCE on the left and right chaotic components was found to be 0.020
and 0.010, respectively.  As expected, each chaotic component has a
different LCE.  This phenomenon was observed in disk galaxies without a
halo component (see Fig. 3 in Papadopoulos \& Caranicolas 2005).  It was
also observed in potentials made up of perturbed harmonic oscillators
(see Saito and Ichimura 1978).  Moreover, the LCE for all orbits,
starting in the chaotic sea of Figure 3, i.e., when $\lambda=0.03, M_n
=400, h_2=500$, was found to be 0.165.  This result is not surprising as
in Figure 3 different chaotic regions have merged.

\sectionb{4}{ORBITS IN THE 3D MODEL}

Let us now turn to the properties of orbits in the 3D potential.
In order to do this, we compute orbits with initial conditions
$(x_0,p_{x0},z_0)$, $y_0=p_{z0}=0$, where $(x_0,p_{x0})$ is a point
inside the limiting curve of the 2D system.  The limiting curve
corresponds to
\begin{equation}
\frac{1}{2}p_x^2+V_T(x)=h_2
\end{equation}
and can be obtained from (7) if we set $y=p_y=0$.  For convenience, we
take $E=h_2$ and the value of $p_{y0}$ is always found from the energy
integral (6).

Our numerical calculations suggest that all orbits in the 3D model,
starting with initial conditions $(x_0,p_{x0})$ on the chaotic zones of
the 2D system of Fig\-ure~1, i.e., when $\lambda=0.015, M_n =400,
E=h_2=500$, and all permissible values of $z_0$, i.e. such as to obtain
real values for $p_y$, are chaotic.  The values of the LCE are different
for the 3D orbits starting in different chaotic zones.  Thus, the value
of the LCE for orbits starting with $(x_0,p_{x0})$ in the chaotic zone
near the center was found to be 0.105.  For the chaotic zone on the
left-hand side of Figure 1, the LCE was found to be 0.015 while for the
chaotic zone on the right-hand side of Figure 1 it was 0.008.  These
results seem to be correct because there is no evidence yet that in 3D
systems with divided phase space a completely connected chaotic
component actually exists (see Cincotta et al. 2006).  On the other
hand, all orbits with initial conditions $(x_0,p_{x0},z_0)$,
$y_0=p_{z0}=0$, with values of $(x_0,p_{x0})$ in the chaotic sea of
Figure 3, were found chaotic for all permissible values of $z_0$.  The
corresponding LCE was found to converge to a common value, equal to
0.124.

In order to have a complete picture of the behavior of orbits in the 3D
system, we have calculated a large number of orbits with initial
conditions, $(x_0,p_{x0},z_0)$, $y_0=p_{z0}=0$ with the values of
$(x_0,p_{x0})$ in the regular regions of Figure 1. It was found that for
the values $|z_0|<4$ all tested orbits were found regular, while when
$|z_0| \geq 4$ the orbits were found chaotic.  The value of the LCE was
found to be\hfil\break

%%%%%%%%%%%%%%%%%%%%%%%%%%  FIGURE 8

\begin{figure}[!th]
\vspace{-4mm}
\resizebox{\hsize}{!}{\rotatebox{0}{\includegraphics*{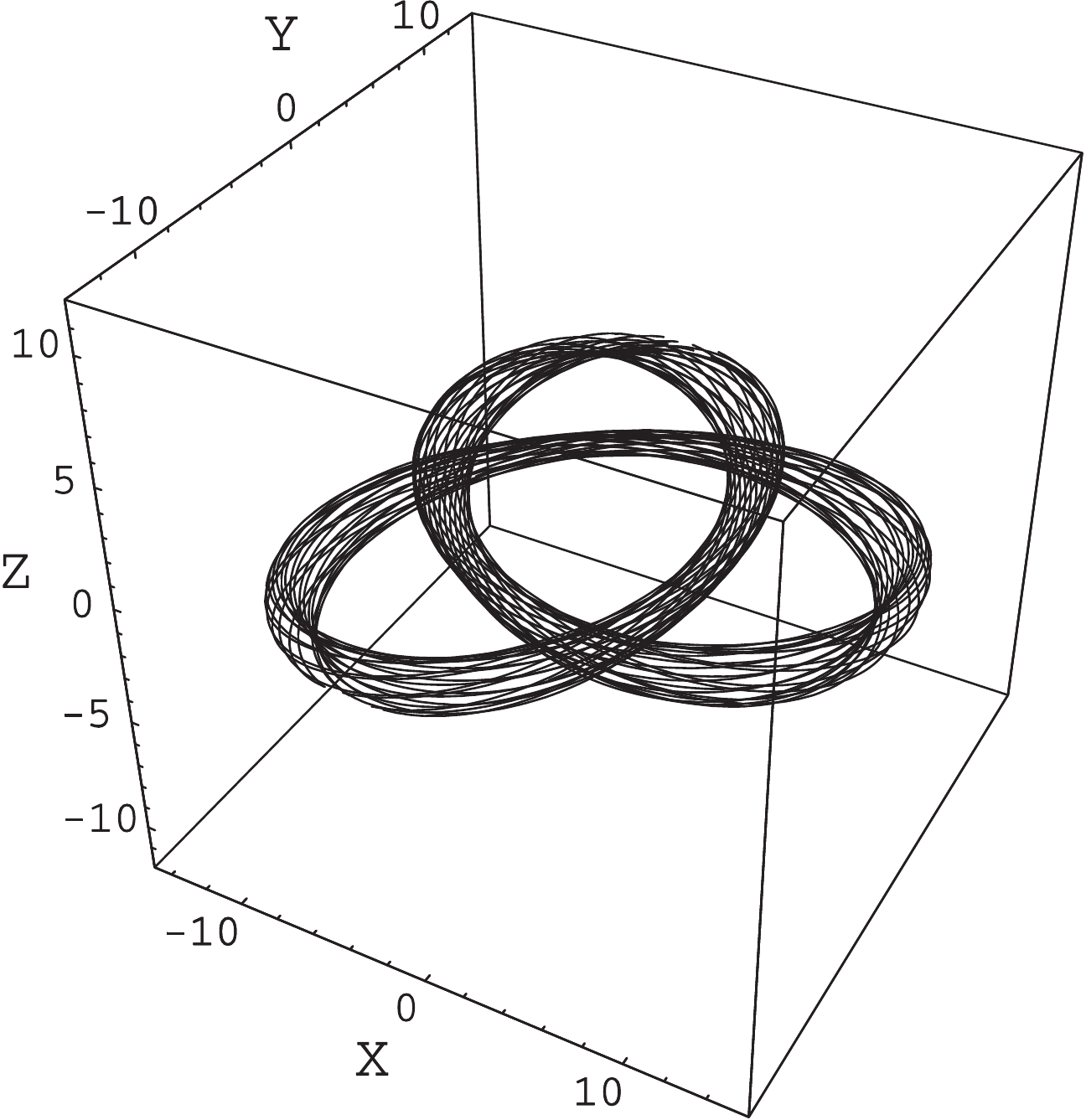}}\hspace{1cm}
                      \rotatebox{0}{\includegraphics*{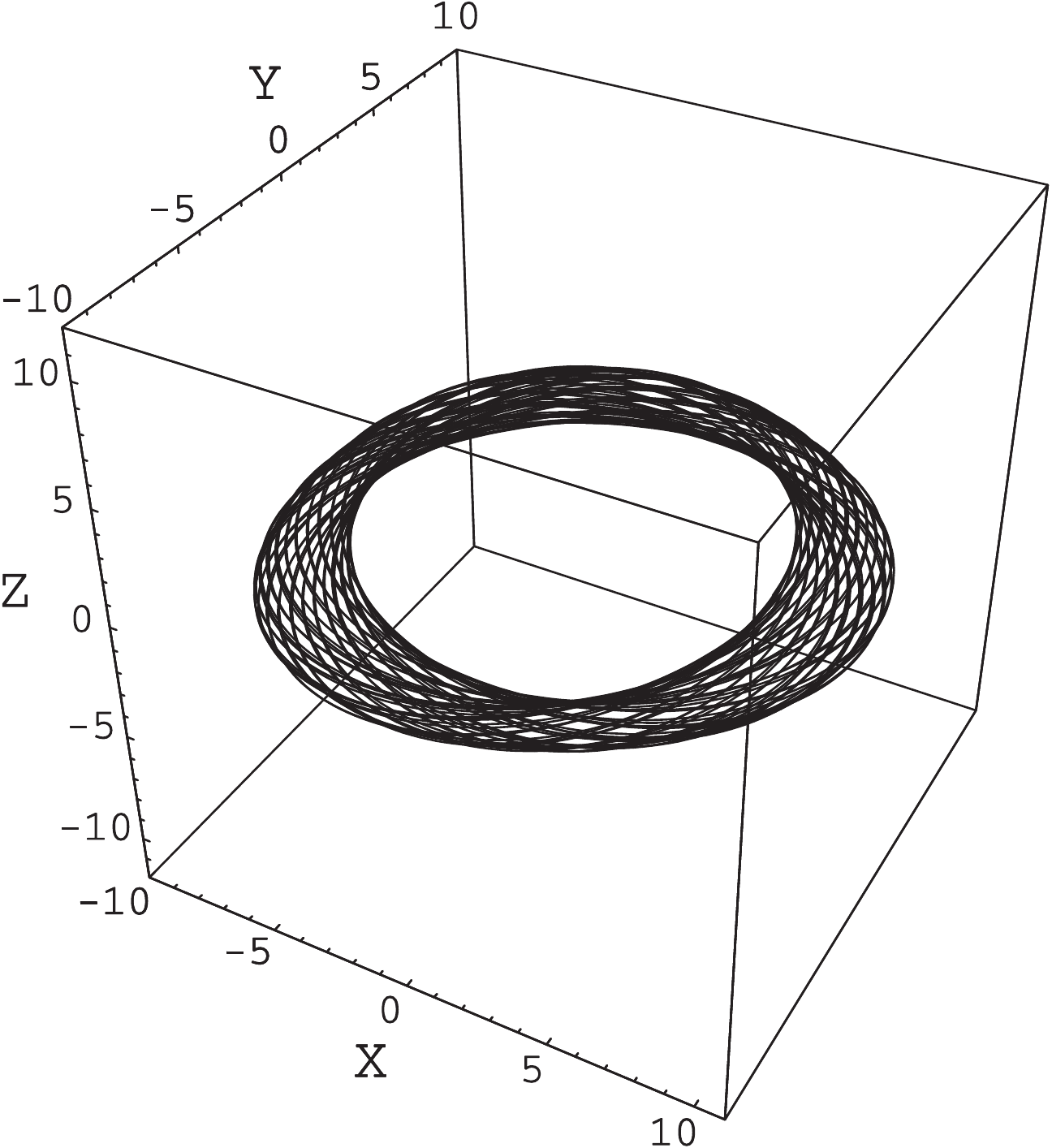}}}
\resizebox{\hsize}{!}{\rotatebox{0}{\includegraphics*{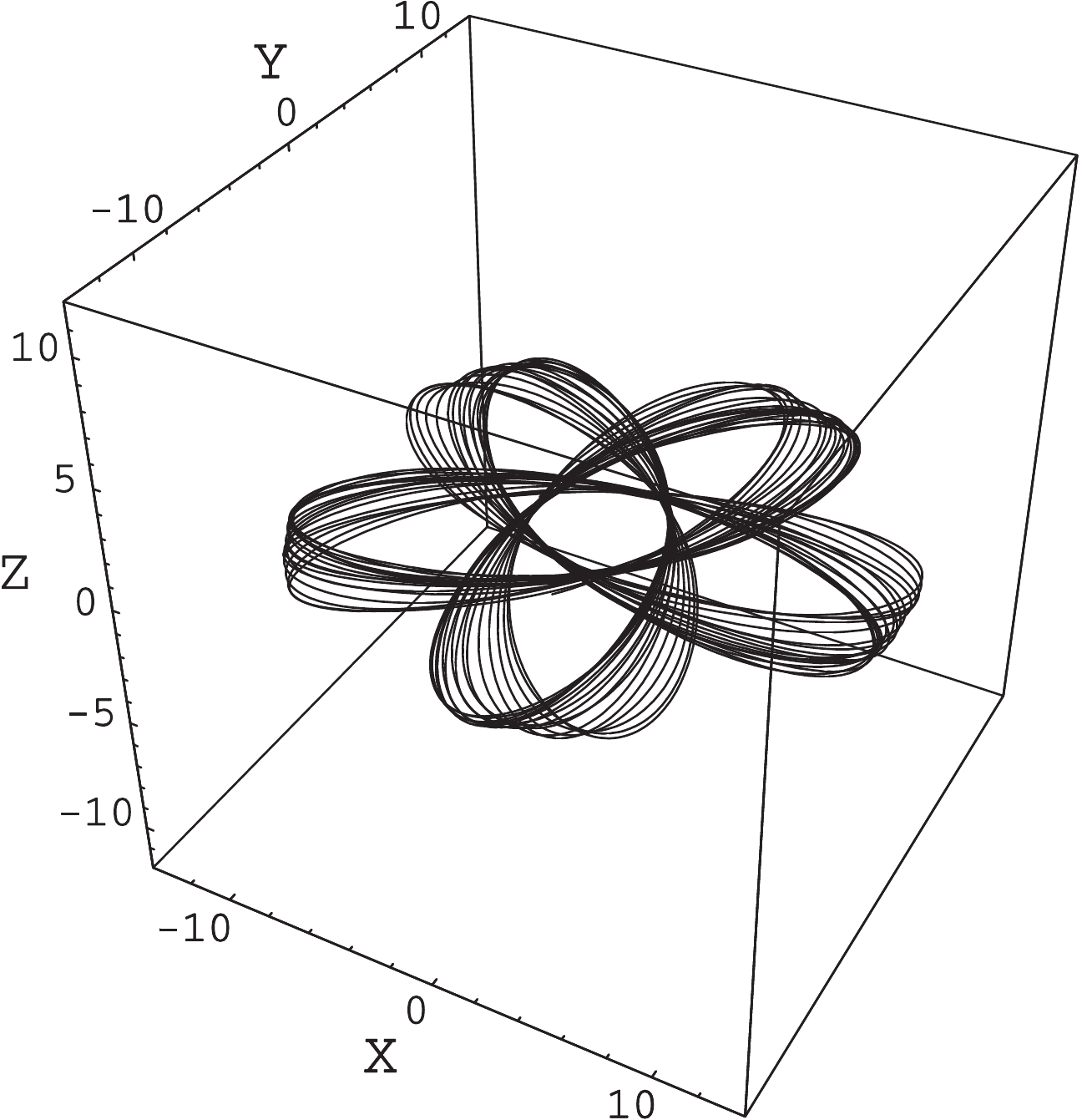}}\hspace{1cm}
                      \rotatebox{0}{\includegraphics*{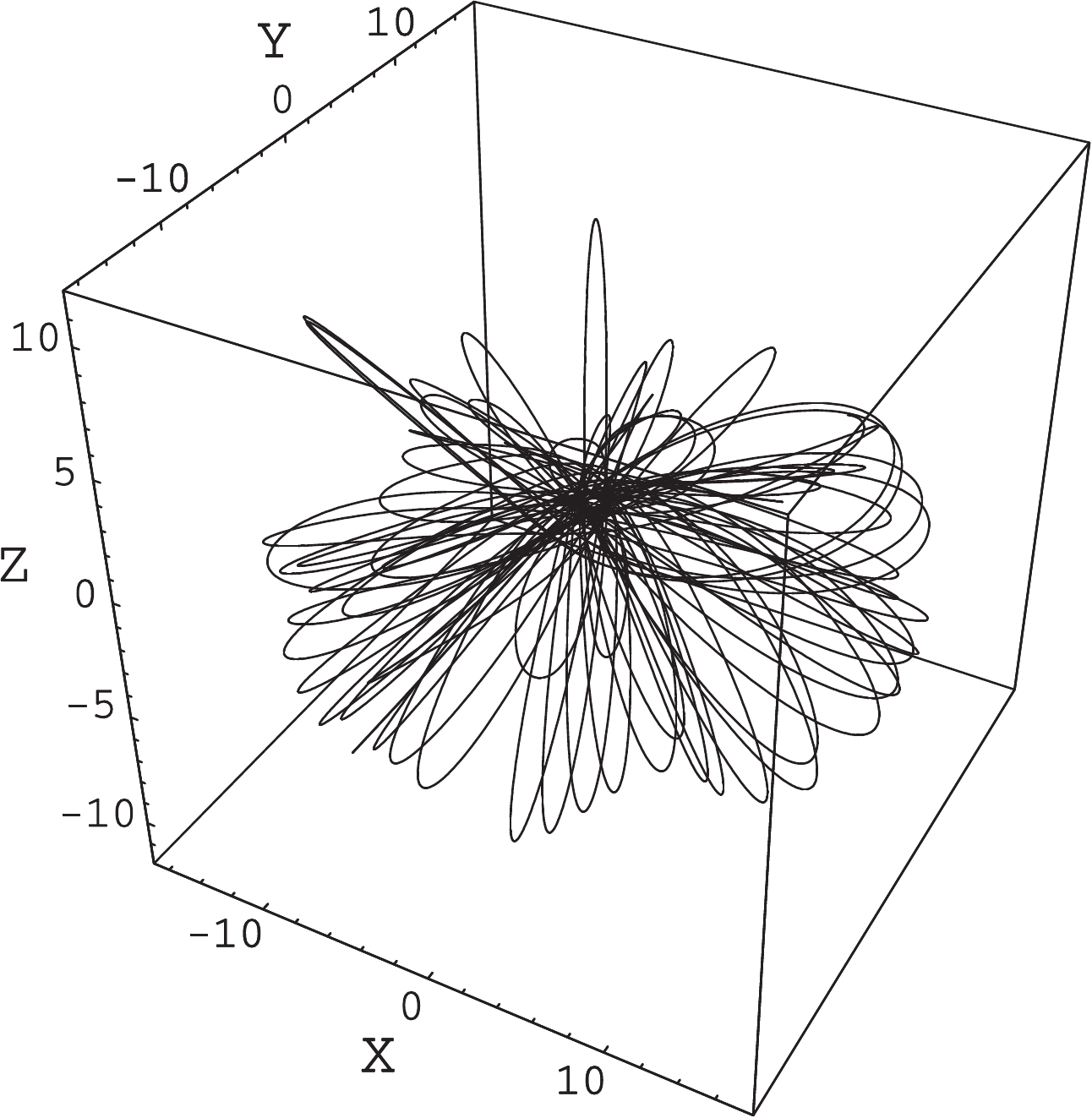}}}
\vskip 0.1cm
\captionb{8}{Panels (a)--(d):  representative orbits in the 3D
potential.  Initial conditions for $x_0,p_{x0}$ as in Figures 5a--5d,
respectively, while $z_0=0.1$ and $E=h_2$ for all orbits.  Note that the
chaotic orbit in Figure 5d is scattered to the halo.}
\end{figure}

\noindent between 0.005--0.008 for orbits starting on the left-hand side
of Figure 1, while for orbits on the right-hand side of Figure 1 the LCE
takes values in a range of 0.010--0.015.

Figures 8\,(a--d) show four representative orbits in the 3D potential.
In order to better visualize the behavior of the 3D orbits, we have used
the same initial conditions of the orbits as in Figures 5\,(a--d) for
the 2D system, with the same value of energy, $E= h_2$.  The value of
$z_0$ was taken to be equal to 0.1 for all orbits.  It is seen that all
orbits retain their regular or chaotic character.  Furthermore, one can
see only chaotic orbits approach the dense nucleus while regular orbits
do not.  All orbits were calculated for a time period of 100 time units.

%%%%%%%%%%%%%%%%%%%%%%%%%%%%  FIGURE 9

\begin{figure}[!th]
\resizebox{\hsize}{!}{\rotatebox{0}{\includegraphics*{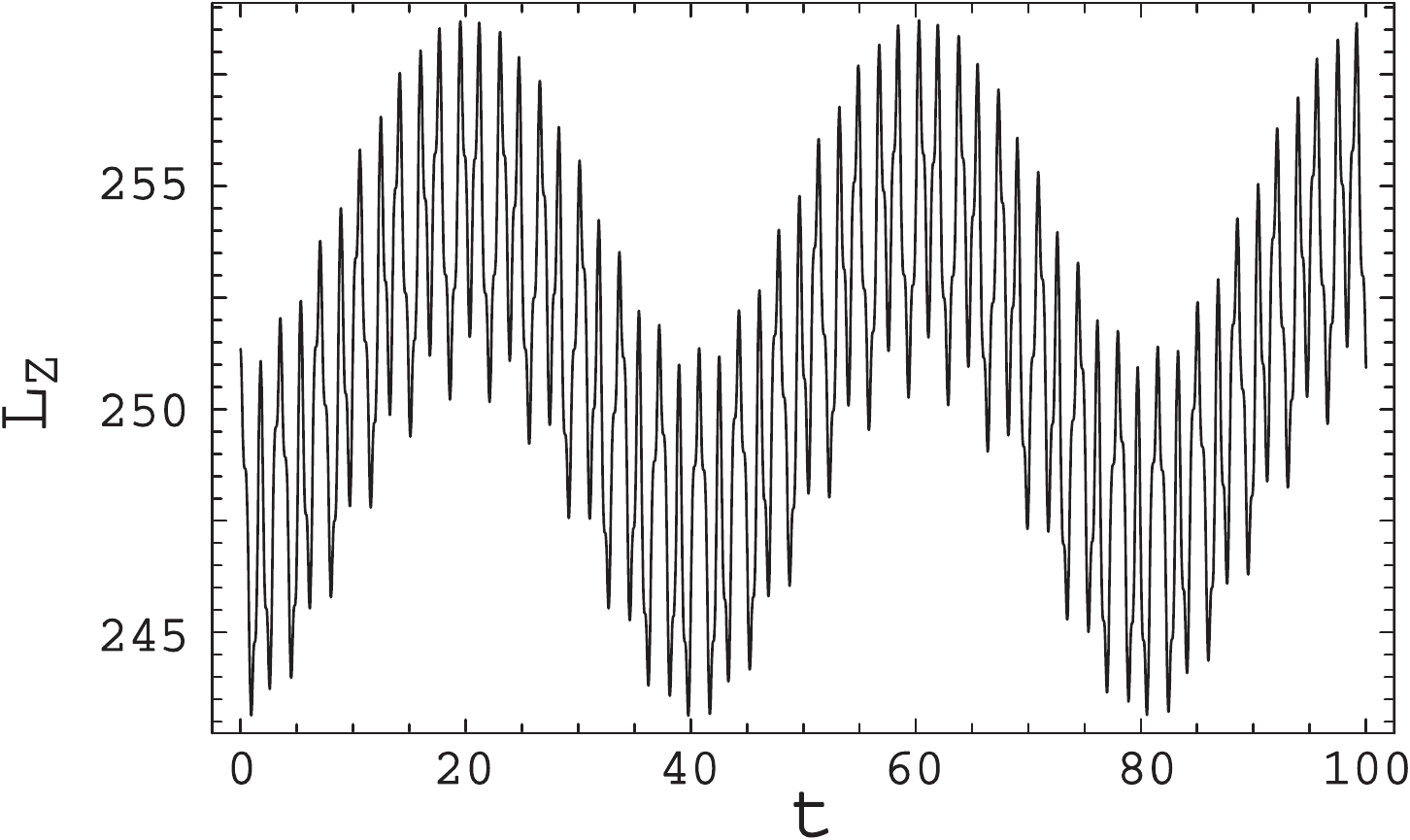}}\hspace{1cm}
                      \rotatebox{0}{\includegraphics*{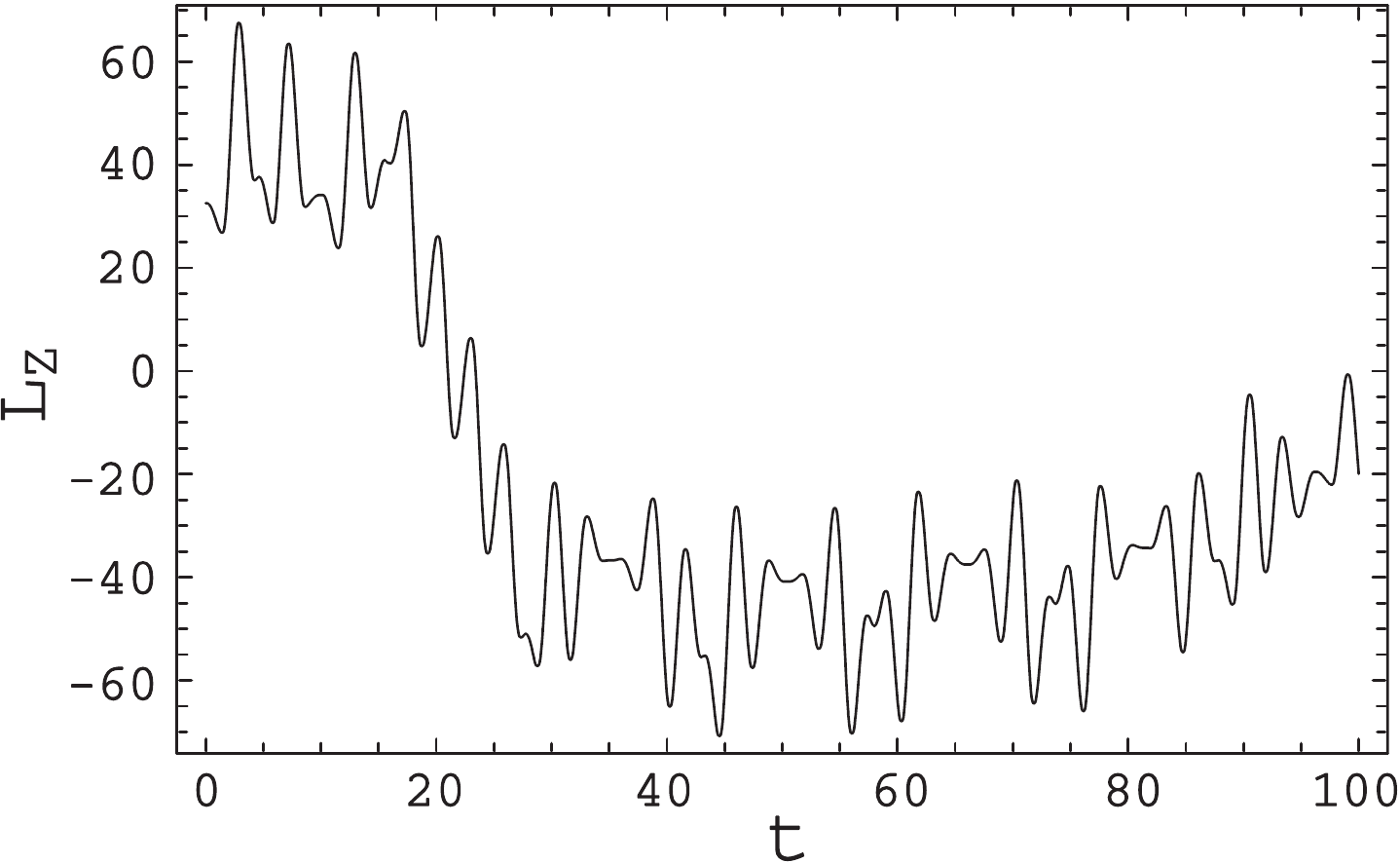}}}
\vskip 0.1cm
\captionb{9}{Evolution of the angular momentum vs. time: the left panel
is for the quasi-periodic orbit of Figure 8b and the right panel
is for the chaotic orbit of Figure 8d.}
\end{figure}

It is also interesting to note that, from all the above four orbits
starting near the galactic plane, only that approaching the nucleus is
deflected to the halo.  Therefore, we can say that orbits moving near
the galactic plane, when approaching the dense nucleus, are scattered to
higher $z$.  This behavior is similar to that observed about twenty
years ago by Caranicolas \& Inannen (1991) in axially symmetrical disk
galaxies, hosting dense nuclei.  Today it is well known that
for this scattering is responsible the strong $F_z$ nuclear force, that
acts upon the low angular momentum stars (see also Caranicolas \&
Papadopoulos 2003).

We would also like to remind the reader that, in any case, the orbits
that are scattered to the halo are orbits of low angular momentum.  As
here the $L_z$ component of the angular momentum is not conserved, we
can compute its mean value
\begin{equation}
<L_z>=\frac{1}{n}\displaystyle\sum_{k=1}^{n}L_{zk}.
\end{equation}

Figure 9a shows the evolution of angular momentum vs. time for the orbit
shown in Figure 8b.  As expected, one can observe a quasi periodic curve
while $<L_z>=251$.  Figure 9b is similar to Figure 9a but for the
chaotic orbit of Figure 8d.  Here one can see an asymmetric curve and
$<L_z>=-27.1$.  Both orbits were calculated for a time period of 100
time units and with $n=10^4$.

\sectionb{5}{DISCUSSION}

It is well known that the evolution of galaxies is closely associated to
the shape and the structure of the host halos.  Today astronomers
believe that dark halos may have a variety of shapes and they usually
show asymmetries (see, e.g., Dinshaw et al. 1998; Oppenheimer et al.
2001; McLin et al. 2002; Penton et al. 2002; Steidel et al. 2002).

In the present paper, the behavior of orbits in a galaxy with an
asymmetric dark halo component was investigated.  In order to keep
things simple, we have chosen a nearly spherical dark halo component
with a small deviation from spherical symmetry, introduced by the term
$-\lambda x^3$.  The results of this work strongly suggest, that small
asymmetries in the halo may play an important role on the regular or
chaotic nature of orbits.

We have first studied the 2D system using the classical method of the
Poincar\'e phase plane.  Our numerical experiments show that there are
three distinct areas of chaotic motion when $\lambda=0.015$ and $h_2$ =
500:  a chaotic layer in the central region and two additional chaotic
regions, which appear near the unstable periodic orbits produced by the
secondary resonances.  As expected, the LCE was found to be different in
each chaotic region.  On the other hand, when $\lambda=0.03$ and $h_2$ =
500, the LCE was found to converge to a unique value when the three
chaotic regions merge to form a large chaotic sea.

Using the results of the 2D system, we have investigated the 3D
potential.  We have used initial conditions $(x_0,p_{x0},z_0)$, where
$(x_0,p_{x0})$ was a point on the phase plane.  It was found, that all
orbits starting in the chaotic regions on the phase plane, produce
chaotic orbits for all values of $z_0$.  Note that in all cases the
value of $p_y$ was found from the energy integral, while $y=p_z=0$.  Our
numerical results indicate that the three chaotic components, found in
the 2D system, exist also in the 3D potential and have also different
values of the LCE when $\lambda=0.015$.  Furthermore, when
$\lambda=0.03$ and the three chaotic components merge, as it was
indicated in the phase plane of the 2D system, the LCE of the 3D system
seems to converge to a unique value.  This result is in agreement with
the outcomes for the 3D systems obtained by Cincotta et al.  (2006).

It was also found that orbits starting in the regular parts of the 2D
system display chaotic motion only when $|z_0| \geq 4$.  Orbits with
low values of $<L_z>$, moving near the galactic plane, when approaching
the dense and massive nucleus, are scattered to the halo displaying
chaotic motion.  This behavior was also observed in axially symmetrical
disk galaxies with dense massive nuclei (see also Caranicolas \&
Papadopoulos 2003 and references therein). Thus, one can conclude that
low angular momentum stars are on chaotic orbits in axially symmetrical,
as well as in triaxial galaxies, hosting massive and dense nuclei.

\vskip3mm

ACKNOWLEDGMENT. The authors would like to thank the referee Peeter
Tenjes for his useful suggestions and comments.

\References

\refb Binney J., Tremaine S. 2008, {\it Galactic Dynamics}, Princeton
Series in Astrophysics, 2nd edition

\refb Caranicolas N. D., Innanen K. A. 1991, AJ, 102, 1343

\refb Caranicolas N. D., Papadopoulos N. 2003, A\&A, 399, 957

\refb Caranicolas N. D., Zotos E. E. 2009, Astron. Nachr., in press

\refb Cincotta P. M., Giordano C. M. et al. 2006, A\&A, 455, 499

\refb Dinshaw N., Foltz C. B. et al. 1998, ApJ, 494, 567

\refb Kunihito I., Takahiro T. et al. 2000, ApJ, 528, 51

\refb Lichtenberg A. J., Lieberman M. A. 1992, {\it Regular and Chaotic
Dynamics}, Springer, 2nd edition

\refb Miyamoto M., Nagai R. 1975, PASJ, 27, 253

\refb McLin K., Stocke J. T. et al. 2002, ApJ, 574, L115

\refb Olling R. P., Merrifield M. R. 2000, MNRAS, 311, 361

\refb Oppenheimer N. C., Hambly A. P. 2001, Science, 292, 698

\refb Papadopoulos N., Caranicolas N. D. 2005, Baltic Astronomy, 14, 253

\refb Penton S. V., Stocke J. T., Shull J. M. 2002, ApJ, 565, 720

\refb Saito N., Ichimura A. 1979, in {\it Stochastic Behavior in
Classical and Quantum Hamiltonian Systems}, Eds.  G. Casati \& J. Ford,
Springer, p. 137

\refb Steidel C. C., Kollmeier J. A. et al. 2002, ApJ, 570, 526

\refb Xu Y., Deng L. C., Hu J. Y. 2007, MNRAS, 379, 1373

\refb Wechsler R. H., Bullock J. et al. 2002, ApJ, 568, 52

\end{document}